\documentclass[aps,prl,twocolumn,showpacs,groupedaddress]{revtex4}
\usepackage{graphicx}  
\usepackage{dcolumn}   
\usepackage{bm}        
\usepackage{amssymb}   
\begin{document}

\def\ra{{\rightarrow}}
\def\a{{\alpha}}
\def\b{{\beta}}
\def\l{{\lambda}}
\def\eps{{\epsilon}}
\def\T{{\Theta}}
\def\t{{\theta}}
\def\co{{\cal O}}
\def\car{{\cal R}}
\def\caf{{\cal F}}
\def\cs{{\Theta_S}}
\def\pr{{\partial}}
\def\tri{{\triangle}}
\def\na{{\nabla }}
\def\S{{\Sigma}}
\def\s{{\sigma}}
\def\sp{\vspace{.06in}}
\def\hs{\hspace{.25in}}
\def\rs{\vspace{-.08in}}

\newcommand{\be}{\begin{equation}} \newcommand{\ee}{\end{equation}}
\newcommand{\bea}{\begin{eqnarray}}\newcommand{\eea}
{\end{eqnarray}}


\begin{titlepage}

\title{\Large\bf ${\mathbf{{CFT}_6}}$ Bulk/Boundary ${\mathbf{{AdS}^Q_5}}$ Correspondence and Emergent Gravity} 

\author{Supriya Kar, R. Nitish and Deobrat Singh}
\affiliation{Department of Physics \& Astrophysics, University of Delhi, New Delhi 110 007, India}

\begin{abstract}
We revisit a non-perturbation theory of quantum gravity in $1.5$ order underlying an emergent gravitational pair of $(4{\bar 4})$-brane with a renewed interest. In particular the formulation is governed by a geometric torsion ${\cal H}_3$ in second order with an on-shell NS form in first order. Interestingly the gravitational pair is sourced by a Kalb-Ramond two form CFT on a $D_5$-brane in $d$$=$$10$ type IIB superstring theory. We show that a generic form theory containing a CFT sector in $d$$=$$6$ bulk may be described by a boundary ${\rm AdS}_5$ with a quintessence Q. Analysis reveals that the bulk/boundary duality in emergent gravity can be a potential tool to explore the quintessential cosmology.

\end{abstract}

\maketitle
\end{titlepage}

\noindent
{\bf Introduction:} Holographic principle \cite{susskind-hologram}, underlying a correspondence between bulk and boundary dynamics, is believed to unfold some of the non-perturbation (NP) tools and their consequences to a perturbation world. In particular the correspondence between a five dimensional anti de Sitter $({\rm AdS}_5)$ gravity in bulk and a four dimensional conformal field theory $({\rm CFT}_4)$ at the boundary is a remarkable phenomenon \cite{maldacena-AdS,witten-AdS,klebanov-witten}. The novel idea has lead to a NP-theory and has been perceived using a strong-weak coupling duality in superstring theory \cite{ashoke}. In fact the bulk/boundary duality has been explored extensively for diverse phenomena in the last two decades. The spontaneous symmetry breaking phenomenon at the event horizon of a black hole has turned out to be exciting \cite{hartnoll,gubser-SSB}. It is believed to lead to an unified description under the NP-quantum theory of gravity. Pioneering attempt to obtain the quantum theory of gravity using gauge theoretic tool(s) have also been explored \cite{wilczek-PRL,wilczek-Unification}.

\sp
\noindent
In the context Dirichlet $(D)$ brane is known to be a NP-dynamical object in $d$$=$$10$ type II superstring theories \cite{polchinski}. A  $D_p$-brane propagation describes a $(p+1)$-dimensional hyper-surface. It can be sourced by a non-linear $U(1)$ boundary dynamics of an open string in presence of a constant Neveu-Schwarz (NS) two form background \cite{seiberg-witten} and hence its dynamics is approximated by the Dirac-Born-Infeld (DBI) action. Mathematical difficulties donot allow an arbitrary NS form on a $D_p$-brane. Nevertheless, the NS form dynamics is known to incorporate a torsion in the string effective action underlying the string world-sheet conformal symmetry. Though certain aspect of a torsion in (super)string theory has been investigated in the past \cite{candelas-HS,freed}, its connection to a $D_p$-brane has not been explored to its strength.

\sp
\noindent
In the recent past author(s) in collaborations, have addressed the NS form gauge dynamics in an emergent scenario on a fat $4$-brane  \cite{abhishek-JHEP,abhishek-PRD,abhishek-NPB-P,sunita-NPB,sunita-IJMPA,priyabrat-EPJC,priyabrat-IJMPA,priyabrat-IJIRSET,deobrat-IJIRSET} and subsequently underlying a ${\rm CFT}_6$ on a fat $5$-brane \cite{richa-IJMPD,deobrat-springer}. In fact, a fat $(p+1)$-brane is governed by a NS form dynamics in an emergent perturbation theory in first order. Investigation has revealed that the emergent phenomenon is described on a gravitational pair of $(p{\bar p})$-brane for $p\le 8$ and incorporates a dynamical NP-correction precisely described by a geometric torsion ${\cal H}_3$ in second order.
Thus a NP-theory of quantum gravity has been shown to be governed by the ${\cal H}_3$ in $1.5$ order for an on-shell NS form \cite{kar-PRD-R}. It may be viewed with the exchange of closed string(s) in between a $D_p$-brane and an anti $D_p$-brane in type II superstring theories. It has been shown that the emergent semi-classical geometries receive a NP-correction sourced by a lower dimensional $D_p$-brane underlying a geometric torsion theory. Recently an intrinsic aspect of NP-tool has been explored to generate mass for the gauge field and NS form in a perturbation gauge theory \cite{kar-nitish}. 

\sp
\noindent
In the article we explore an important aspect of NP-tool leading to two equivalent dynamical description in terms of bulk ${\rm CFT}_6$ and boundary ${AdS}^Q_5$ underlying a fat $5$-brane. Unlike to ${\rm AdS}_5/{\rm CFT}_4$ correspondence \cite{maldacena-AdS,witten-AdS}, the ${\rm CFT}_6$ in bulk/boundary ${\rm AdS}^Q_5$ underlies an emergent NS form theory in first order. The proposed correspondence is a generic feature between a NS form dynamics on a fat $(p+1)$-brane and an emergent metric dynamics on a $p$-brane in a NP-formulation.

\sp
\noindent
{\bf ${\mathbf{D_p}}$-brane and NS form:}
We begin with a (bosonic) open string world-sheet dynamics in presence of the massless closed string backgrounds: metric $g_{\mu\nu}(X)$, Neveu-Schwarz (NS) two form $B_2^{(NS)}(X)$, and dilatonic scalar $\Phi(X)$. The non-linear sigma model action underlie an open string $X^{\mu}(\sigma,\tau)$ in the world-sheet bulk. For a constant $\Phi$, the world-sheet action may be given by 
\rs
\be
S=-T\int d^2\sigma\Big (\sqrt{-h}h^{ab}g_{\mu\nu} + \epsilon^{ab}{\bar F}^{nl}_{\mu\nu}\Big )\partial_aX^{\mu}\partial_b X^{\nu}\ ,\label{WS-1}
\ee

\rs
\noindent
where $T=(2\pi\alpha')^{-1}$ is the fundamental string tension, $h$ is the determinant of the world-sheet metric $h_{ab}$. The non-linear field strength ${\bar F}^{nl}_{\mu\nu}= (2\pi\alpha')F^{nl}_{\mu\nu}$ remains $U(1)$ gauge invariant in a combination of transformations for the gauge fields $A_{\mu}$ and $B_{\mu\nu}^{(NS)}$. Explicitly:
\rs
\be
{\bar F}^{nl}_{\mu\nu}=B_{\mu\nu}^{(NS)}+ {\bar F}_{\mu\nu}\ ,{\rm where}\;\ F_{\mu\nu}=\nabla_{\mu}A_{\nu}- \nabla_{\nu}A_{\mu}\ .
\label{WS-2}
\ee
The linear field strength $F_{\mu\nu}$ is uniform, which is indeed an electromagnetic field. However the modified $F^{nl}_{\mu\nu}$ turns out to be non-uniform in the bulk in presence of a NS form. Under the $U(1)$ gauge transformation of forms, 
$\delta A_{\mu}$$=$${\nabla}_{\mu}\epsilon$ and $\delta B_{\mu\nu}^{(NS)}$$=$$\big (\nabla_{\mu}{\cal E}_{\nu}$$-$$\nabla_{\nu}{\cal E}_{\mu}\big )$, the respective field strengths $F_{\mu\nu}$ and $H_{\mu\nu\lambda}^{(NS)}$$=$$3\nabla_{[\mu}B_{\nu\lambda]}^{(NS)}$ remain invariant. They 
contribute to the string effective action, which is obtained using the conformally invariant world-sheet $i.e.\ \beta$-function(s) $=$$0$.

\sp
\noindent
For a constant NS form in the open string bulk, $H_{\mu\nu\lambda}^{(NS)}$$=0$ in the string effective action. Then, the field strength $F_{\mu\nu}^{nl}$ turns out to be a constant. The second term in the bulk (\ref{WS-1}) is re-expressed as a boundary integral, which perceives the propagation of $D_p$-brane with $(25$$-$$p)$ Dirichlet conditions. The DBI action underlying a $D_P$-brane is given by
\rs
\be
S_{\rm DBI}= {{-1}\over{g_s (2\pi)^p(\alpha')^{{(p+1)}/2}}}\int d^{p+1}x\ \sqrt{g_{\mu\nu} + {\bar{F}}^{nl}_{\mu\nu}}\ .\label{DBI-1}
\ee

\rs
\noindent
It shows that a $D_p$-brane is precisely governed by $A_{\mu}$, while its global property is modified by a constant NS form. Since a global NS form cannot be gauge away, it modifies a point charge to a non-linear charge and hence $F_2$$\rightarrow$$F^{nl}_2$. However the $U(1)$ gauge invariant $F_2^{nl}$$\rightarrow$$B_2^{(NS)}$ as $F_2$ can be gauged away.

\sp
\noindent
Furthermore a $D_p$-brane is flat, $i.e.$ defined with a constant metric $g_{\mu\nu}$, as the closed strings are tangential to its world-volume. A  constant NS form modifies the constant $g_{\mu\nu}$ and leads to an open string metric 
${\tilde{G}}_{\mu\nu}$$=$$\big ( g_{\mu\nu}$$-$$B_{\mu}^{(NS)\lambda}B^{(NS)}_{\lambda\nu}\big )$ on a $D_p$-brane \cite{seiberg-witten}. Arguably the open string metric defines a nontrivial potential generated by a global mode of NS form on a $D_p$-brane. In the recent past, the metric potential is explored to obtain various near horizon black hole geometries \cite{gibbons,mars,ishibashi,kar-majumdar,kar-PRD,kar-JHEP,liu,zhang2}.

\sp
\noindent
{\bf ${\mathbf{D_4}}$-brane and KR form:}
Generically the DBI action can be re-expressed in terms of a higher $(p\ge 2)$ form $U(1)$ gauge theory on a $D_{(p+2)}$-brane. For simplicity $p$$=$$2$ has been explored in the recent past
\cite{abhishek-JHEP,abhishek-PRD,abhishek-NPB-P,sunita-NPB,sunita-IJMPA,priyabrat-EPJC,priyabrat-IJMPA,priyabrat-IJIRSET,deobrat-IJIRSET,richa-IJMPD,deobrat-springer} to explain diverse phenomena underlying a semi-classical emergent theory of metric in a NP-formulation of quantum gravity on a gravitational pair. 

\sp
\noindent
Very recently the NP-formulation was further reviewed with a renewed interest to obtain a dynamical geometric torsion correction ${\cal F}_4=\big (d{\cal H}_3-{\cal H}_3\wedge {\cal F}_1\big )$ in second order to an emergent metric \cite{kar-PRD-R}. An attempt has been made to obtain the $M$-theory in $d$$=$$11$ on a gravitational pair of $(M{\bar M})$-brane from $d$$=$$12$ form theory underlying a fundamental gravitational $3$-brane dynamics. The NP-tool has further been exploited to generate mass for the NS-form on a fat $3$-brane \cite{kar-nitish}. It was argued that an axionic scalar dynamics ${\cal F}_1$$=$$d\psi$ underlying a NP-correction is absorbed to generate a massive NS form in a perturbative vacuum. Interestingly $\psi$ plays the role of a goldstone boson in perturbation theory. It transforms a flat $D_4$-brane to a fat $4$-brane.

\sp
\noindent
In fact, the non-linear $U(1)$ gauge dynamics on a $D_4$-brane has equivalently been realized with a linear $U(1)$ symmetry \cite{seiberg-witten}. Interestingly the gauge theory may also be re-expressed with a Kalb-Ramond (KR) two form dynamics using the Poincare duality.
Then, the non-linear gauge dynamics of a $D_4$-brane, defined with a constant NS form and a local KR form, in presence of a background (open string) metric may be given by
\rs
\be
S={{-1}\over{(8\pi^3g_s){\alpha'}^{3/2}}}\int d^5x {\sqrt{-{\tilde G}^{({\rm NS})}}}\ H_{\mu\nu\lambda}H^{\mu\nu\lambda}\ ,\label{p-1}
\ee

\rs
\noindent
where $H_{\mu\nu\lambda}$$=$$3\nabla_{[\mu}B_{\nu\lambda]}^{(KR)}$ is sourced by a string charge.

\sp
\noindent
{\bf Fat ${\mathbf{4}}$-brane and NS form:}
In particular a KR quantum, in the world-volume gauge theory on a $D_4$-brane, has been argued to vacuum create a stringy pair across an event horizon of a background black hole underlying the fundamental principle of Schwinger pair production mechanism \cite{schwinger}. The NP-tool has further been explored to explain the Hawking radiation phenomenon \cite{hawking} at the event horizon of a black hole. The mechanism has been explored for an open string pair production by an electro-magnetic field \cite{bachas-porrati} and for a vacuum pair of $(D{\bar D})_9$ at the cosmological horizon \cite{majumdar-davis}.

\sp
\noindent
In the context the absorption of KR quanta leading to an emergent stringy pair has been realized geometrically, when $H_{\mu\nu\lambda}$ is exploited as a torsion connection \cite{abhishek-JHEP,abhishek-PRD,abhishek-NPB-P}. A torsion modifies the covariant derivative operator ${\nabla}_{\mu}$ on a $D_4$-brane to ${\cal D}_{\mu}$ on an emergent fat $4$-brane. The modified covariant derivative operation on a NS form turns out to be significant and is given by
\rs
\be
{\cal D}_{\lambda}B^{(NS)}_{\mu\nu}={1\over2}\left ({H_{\lambda\mu}{}}^{\rho}B^{(NS)}_{\nu\rho}+ {H_{\lambda\nu}{}}^{\rho}B^{(NS)}_{\rho\mu}\right )\ ,\label{gtorsion-1}
\ee

\rs
\noindent
where $\nabla_{\mu}B^{(NS)}_{\mu\nu}$$=$$0$ as the NS form is covariantly constant on a $D_4$-brane.  An emergent fat brane evolves with a dynamical NS field and is obtained at the expense of the KR field dynamics on a $D_4$-brane. It has been shown that the emergent dynamical NS form defines a geometric torsion: ${\cal H}_{\mu\nu\lambda}$$=$$3{\cal D}_{[\mu}B^{(NS)}_{\nu\lambda]}$. Thus a constant NS form on a $D_4$-brane has lead to a nontrivial ${\cal H}_3$ on an emergent fat $4$-brane. It may well be understood via a coupling of a closed string to a $D_4$-brane. It shows that a fat brane takes account for the quantum gravity underlying a NP-formulation. The NS form perturbation theory on a fat $4$-brane may explicitly be viewed in terms of the KR form gauge theory on a $D_4$-brane. It is given by
\rs
\be
{\cal H}_{\mu\nu\lambda}=H_{\mu\nu\rho}B_{\lambda}^{(NS)\rho}+ H_{\mu\nu\alpha}B_{\rho}^{(NS)\alpha} B_{\lambda}^{(NS)\rho}+\dots\label{gtorsion-2}
\ee

\rs
\noindent
A constant NS form turns out to be a perturbation parameter in the series expansion. Under an iterative correction: 
$H_3\rightarrow {\cal H}_3$, the perturbation series (\ref{gtorsion-2}) in $B^{(NS)}_2$ may equivalently be described as a NP-theory of a geometric torsion. The geometric torsion ${\cal H}_3$ retains the $U(1)$ gauge invariance under NS form transformation in an emergent perturbation theory defined with a modified covariant derivative ${\cal D}_{\mu}$. However, the gauge invariance in a perturbation series (\ref{gtorsion-2}) is apparently broken. Nevertheless, the gauge invariance has explicitly been restored in the NS form theory \cite{abhishek-JHEP,priyabrat-EPJC} in presence of a symmetric fluctuation: $f_{\mu\nu}$$=$${\bar{\cal H}}_{\mu\alpha\beta}{{\cal H}^{\alpha\beta}{}}_{\nu}$, where ${\bar{\cal H}}_3=(2\pi\alpha'){\cal H}_3$. It has lead to a dynamical (emergent) metric: 
\rs
\be 
G_{\mu\nu}=\Big ( g_{\mu\nu}-B_{\mu}^{(NS)\lambda}B^{(NS)}_{\lambda\nu} + {\bar{\cal H}}_{\mu\lambda\rho}{{\cal H}^{\lambda\rho}}_{\nu}\Big )\ . \label{gauge-metric}
\ee

\rs
\noindent
The local degrees of the metric on an emergent $3$-brane is sourced by a dynamical NS form on a fat $4$-brane in first order. Thus the emergent $(3{\bar 3})$-brane is identified as a gravitational pair. It ensures a fact that GTR emerges via a NP-tool in one higher dimension, $i.e.$ in $d$$=$$5$. A  geometric torsion in eq(\ref{gauge-metric}) incorporates an intrinsic angular momentum and naturally governs the Kerr family of black holes as a vacuum geometry \cite{sunita-NPB,sunita-IJMPA}.

\sp
\noindent
{\bf Gravitational pair and Higher-essence:} The $U(1)$ gauge invariant ${\cal H}_3$ on a fat $(p+1)$-brane leads to an emergent metric dynamics on 
a $p$-bane within a pair. The fact may be viewed as a generalization of the open string metric on a $D_p$-brane \cite{seiberg-witten}. At a first sight, the emergent gravitational $4$-brane (metric) dynamics sourced by $f_{\mu\nu}$ imposes $15$-conditions on a NS form on a fat $5$-brane. However they ensure an on-shell NS form, $i.e.\ {\cal D}^{\lambda}{\cal H}_{\lambda\mu\nu}$$=$$0$, in $1.5$ order and hence describes a propagating ${\cal H}_3$ in a NP-theory \cite{kar-PRD-R}. Thus an emergent two form curvature ${\cal K}_{\mu\nu}$$=$${\cal D}^{\lambda}{\cal H}_{\lambda\nu\mu}$ becomes trivial which results in a four form ${\cal F}_4$ correction in second order.

\sp
\noindent
A priori the KR form $U(1)$ gauge theory in $d$$=$$6$ may be identified as the bulk. A fat $5$-brane dynamics has been realized as an emergent metric dynamics on a vacuum created pair of $(4{\bar 4})$-brane underlying a NP-formulation. Thus an emergent $(4{\bar 4})$-brane has been identified as a gravitational pair. A gravitational $4$-brane is governed by the metric dynamics in $d$$=$$5$ and hence describes the Riemannian geometry. However the presence of a gravitational ${\bar 4}$-brane within a pair ensures a higher-essence (or hidden-essence or higher-dimensional) scalar (HS) to a $4$-brane. Alternately, the HS may be identified with an extra dimension transverse to a gravitational $4$-brane as the ${\bar 4}$-brane is hidden across an event horizon \cite{richa-IJMPD,deobrat-springer}.

\sp
\noindent
{\bf ${\rm\mathbf{{CFT}_6}}$ Bulk/Boundary ${\rm\mathbf{{AdS}_5^Q}}$:}
In the context the perspectives of CFT, underlying a KR form $U(1)$ gauge theory, on a $D_5$-brane may play an important role. A traceless energy-momentum tensor for the KR form ensures the conformal symmetry in the classical theory. The conformal anomaly can be set to vanish in a quantum field theory (QFT). This in turn describes a CFT for a KR form in $d$$=$$6$. It has been shown that the gauge theoretic vacuum may equivalently be described by a massless NS form in an emergent $d$$=$$6$ perturbation theory on a fat $5$-brane \cite{richa-IJMPD}. A pair-symmetric emergent curvature tensor of order four has been shown to be sourced by a NS form with $6$ local degrees in first order. They underlie an emergent NP-theory in $d$$=$$5$ on a fat $4$-brane. The NP-correction ${\hat{\cal F}}_4=\big (d{\hat{\cal H}}_3$$-$${\hat{\cal H}}_3\wedge {\hat F}_1\big )$ in $d$$=$$6$ incorporates four local degrees. A plausible ${\hat F}_5$$=$$d{\hat B}_4$ is Poincare dual to ${\hat F}_1=d{\hat\psi}$ and possesses one local degree \cite{kar-PRD-R}. Thus a $B_p$ form theory in $d$$=$$6$ is described by $11$-local degrees of freedom underlying $p=(2,3,4)$, which are respectively Poincare dual to the $p'$-forms for $p'=(2,1,0)$. Generically a form theory in $d$$=$$6$ may be given by
\be
S=-{1\over{12{\hat\kappa}^4}}\int_{bulk} d^6x\ {\sqrt{-{\hat g}}}\ \Big ({\hat{\cal H}}_3^2 -{{{\hat\kappa}^2}\over{4}} {\hat{\cal F}}_4^2 -{1\over{20}} {\hat F}_5^2\Big )\ ,\label{form-1}
\ee
where ${\hat\kappa}$$=$$\sqrt{2\pi\alpha'}$. The NP-correspondence between bulk CFT sector and boundary gravity (BG) in $d$$=$$5$ may be invoked in addition to a dimensional reduction on $S^1$ for the remaining curvature sector described by ${\hat{\cal F}}_4$ and ${\hat F}_5$. The emergent gravity in $d$$=$$5$ has been obtained on a gravitational $4$-brane within a pair of $(4{\bar 4})$-brane \cite{richa-IJMPD}. 

\sp
\noindent
The curvatures $({\hat{\cal H}}_3, {\hat{\cal F}}_4, {\hat F}_5)$ in the form theory correspond respectively to $({\cal R},d\phi_{HS})$, $({\cal F}_4, {\cal F}_3)$ and $(\Lambda, F_4)$ in $d$$=$$5$ emergent gravity. A constant five form in the boundary gravity theory signifies a cosmological $\Lambda=b({\cal E}^{\mu\nu\lambda\rho\sigma}F_{\mu\nu\lambda\rho\sigma})=(-120)b^2$ and hence describes an anti-de Sitter (AdS) geometry. 

\begin{figure}[h]
\centering
\mbox{\includegraphics[width=1.1\linewidth,height=0.3\textheight]{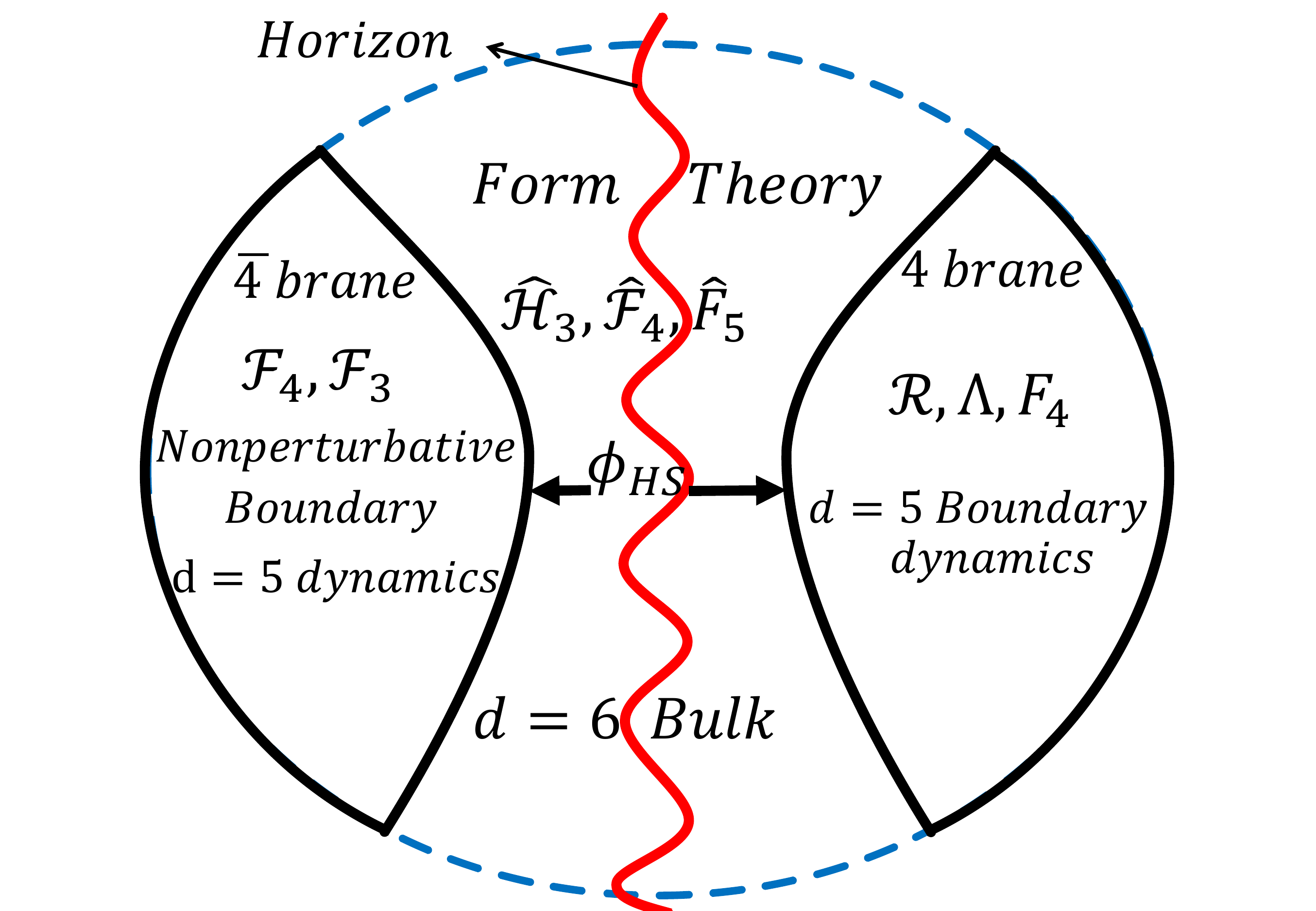}}

\vspace{-.1in}
\noindent
\caption{\it Schematic diagram for ${\rm CFT}_6$ bulk/boundary ${\rm AdS}^Q_5$ correspondence in NP-theory of quantum gravity. The NS form in bulk ${\rm CFT}_6$ sector corresponds to the ${\rm AdS}_5$ boundary on an emergent gravitational $4$-brane and the NP-boundary dynamics on ${\bar 4}$-brane across a horizon. The $4$-brane and ${\bar 4}$-brane may describe the quintessence cosmology and the nuclear interactions respectively. A dynamical HS inbetween the emergent pair presumably reveals a ``gravito-nuclear'' phase.}
\end{figure}

\vspace{-.1in}
\noindent
In a NP-decoupling limit, the weakly coupled $4$-brane decouples from the strongly coupled ${\bar 4}$-brane. It is schematically shown in FIG-1. In the limit the HS dynamics, ${\cal F}_4$ and ${\cal F}_3$ on ${\bar 4}$-brane decouple. Then the boundary theory, with a coupling $\kappa=\sqrt{2\pi\alpha'}$, governs a gravitational $4$-brane and is given by
\be
S={1\over{\kappa}^3}\int_{BG} d^5x\ {\sqrt{-g}}\ \Big ({\cal R}-\Lambda -{1\over{48}}F_4^2\Big )\ .\label{form-2}
\ee

\rs
\noindent
The $4$-form signifies the presence of an interacting axionic scalar field in a semi-classical theory of gravity. The non-canonical potential function for the axionic field is generated by the gravitational interaction. Interestingly the axionic field may be identified with a quintessence scalar dynamics in $d$$=$$5$ and is hidden to the GTR on an emergent $3$-brane within a pair of $(3{\bar 3})$-brane. 

\sp
\noindent
Furthermore the semi-classical theory is governed by $6$-local degrees. The difference of $5$-local degrees, between the bulk and boundary theories, govern the decoupled NP-dynamics on an anti $4$-brane. The momentum conservation at a pair production vertex ensures that the gravitational $4$-brane and the NP-dynamics on an anti $4$-brane moves away from each other along a generalized coordinate $\phi_{HS}$ across a horizon. 

\sp
\noindent
Apparently the emergent scenario implies a repulsive force hidden between a brane-universe and an anti-brane. Generically the HS field dynamics leading to a repulsive force may intuitively be realized with the coulomb force law between the same Ramond-Ramond (RR) charges from the perspective of an emergent gravitational $4$-brane. In particular a vacuum pair generates an equal and opposite charge on a brane and an anti-brane across a horizon. Arguably a conserved mass, defined with a squared charge, changes its sign under a flip of light-cone at the horizon. It generates a repulsive force between a mass-pair \cite{majumdar-davis,richa-IJMPD}.

\sp
\noindent
{\bf Hidden-essence in dual Riemann tensor:} 
The $d$$=$$5$ semi-classical gravity, underlying a boundary theory, may be re-expressed in terms of left (L) and right (R) duals of the Riemann tensor \cite{kar-PRD-R}. The $Q$-essence coupling is absorbed within the duals and they are: 
\rs
\bea
{\cal R}^{(L)}_{\mu\nu\lambda\rho}&=& (2\pi\alpha')\left ( F_{\mu\nu\alpha\beta}\ {{\cal R}^{\alpha\beta}{}}_{\lambda\rho}\right )\nonumber\\
{\rm and}\qquad {\cal R}^{(R)}_{\mu\nu\lambda\rho}&=& (2\pi\alpha')
\left ({{\cal R}_{\mu\nu}{}}^{\alpha\beta}\ F_{\alpha\beta\lambda\rho}\right )\ .\label{form-4}
\eea

\rs
\noindent
The Riemann duals are checked for their irreducibility. A priori the dual(s) Ricci tensor of order two and a Ricci scalar are given by
\bea
{\cal R}^{(L)}_{\lambda\mu}&=&{\kappa^2}\big ( {F_{\lambda\rho}{}}^{\alpha\beta}\ {{\cal R}_{\alpha\beta\mu}{}}^{\rho}\big )=
{\cal R}^{(R)}_{\mu\lambda}=0\nonumber\\
{\cal R}^{(L)}&=&{\cal R}^{(R)}= \kappa^2\big ( {\cal R}_{\mu\nu\lambda\rho}\ F^{\mu\nu\lambda\rho}\big )=0\ .\label{form-5}
\eea
The cyclicity property of the Riemann tensor ensures the irreducibility of ${\cal R}^{(L)}_{\mu\nu\lambda\rho}$ and ${\cal R}^{(R)}_{\mu\nu\lambda\rho}$.
The semi-classical theory of gravity (\ref{form-2}) may be re-expressed in terms of the dual(s) of Riemann tensor and is given by
\be
S={1\over{\kappa^3}}\int_{BG}{\sqrt{-g}}\Big ({\cal R}^{(L)}_{\mu\nu\lambda\rho}R_{(L)}^{\mu\nu\lambda\rho}+
{\cal R}^{(R)}_{\mu\nu\lambda\rho}R_{(R)}^{\mu\nu\lambda\rho}-\Lambda\Big )\label{form-6}
\ee
Remarkably the geometric duals hide an intrinsic coupling of $F_4={}^{\star}d\psi$ with the Riemann tensor. The coupling ensures 
an additional axionic scalar $\psi$ dynamics in a metric tensor theory underlying the Riemannian geometry in $d$$=$$5$. The dynamical $\psi$-correction  may be interpreted as a quintessence, which turns out to be a hidden correction to the GTR. Quintessence is known as a potential candidate to describe the dark energy in universe and hence the semi-classical theory of gravity (\ref{form-6}) may play a vital role to explore the dark gravity \cite{padmanabhan}. Needless to mention that the boundary gravity dynamics in an emergent pair production scenario differs from the novel idea of Kaluza-Klein compactification due to the underlying ${\rm CFT}_6$ bulk/boundary ${\rm AdS}^Q_5$ correspondence. 

\sp
\noindent
Nevertheless, the dynamical contribution of quintessence becomes significant in $d\ge 5$. Thus five space-time dimensions turn out to be the minimal to explore a (non-supersymmetric) NP-theory. Primarily the dynamical correspondence in $d$$=$$5$ is between a KR form in gauge theory and a NS form in superstring theory. Both of them are two forms and they are different due to their differences in connection. Further investigation reveals that a quintessence is described by an axionic scalar in $d$$=$$5$, while its role (sixth essence) in $d$$=$$6$ is described by a massless gauge field $A_{\mu}$ with four local degrees of freedom. A count for the local degrees of NS form, in an emergent perturbation theory, precisely matches with that of a NP-correction in $d$$=$$7$ as ${\cal H}_3$ is Poincare dual to ${\cal F}_4$. Generically the number of local degrees in a NP-correction is greater than the local degrees in a perturbative emergent theory for $d\ge 8$. An emergent gravitational pair of $(8{\bar 8})$-brane describes a space-filling $D_9$-brane in type IIB superstring theory on $S^1$, which is equivalent to the type IIA superstring on $S^1$. Thus a NP-dominance in a supersymmetric formulation would like to begin with a minimal $d$$=$$10$ underlying a gravitational pair $(8{\bar 8})$-brane. It is in agreement with the strongly coupled NP-regime realized with the ${\rm CFT}_4$ boundary dynamics on a $D_3$-brane underlying the ${\rm AdS}_5$ bulk in $d$$=$$10$ superstring theory. With a subtlety for an extra eleventh (small) dimension within a pair of $(9{\bar 9})$-brane, the supersymmetric NP-formulation of emergent gravity presumably justifies the $d$$=$$11$ non-perturbation $M$-theory.

\sp
\noindent
In the context the bulk/boundary correspondence in an emergent gravity may appear to be surprising when compared with the ${\rm AdS_5}$/${\rm CFT}_4$ 
correspondence \cite{maldacena-AdS,witten-AdS,klebanov-witten}. However the apparent puzzle may be resolved in an emergent gravity, where bulk and boundary theories are governed in a NP-formulation. A subtle comparison with the bulk-gravity/boundary-CFT duality may imply that a NS field or generically a $p$-form for $p\ge2$ is likely to incorporate gravitational effect(s) in the boundary, as they are sourced by a string charge for $p$$=$$2$ and a higher dimensional extended charge for $p$$>$$2$.  Intuitively the boundary dynamics in a form theory, underlying an emergent scenario of pair production, is governed by an induced metric underlying a symmetric $f_{\mu\nu}$. Arguably, $f_{\mu\nu}$ possesses a source in the metric background 
for the (super)string world-sheet. A closed string exchange between an emergent $(p{\bar p})$-brane pair further ensures the metric dynamics underlying the ${\rm CFT}_6/{\rm AdS}^Q_5$ correspondence.

\sp
\noindent
In addition the dynamical aspect of a two form inspires to believe in higher form {\it fundamental} theory in $d$$=$$12$. It has been shown to describe an emergent $M$-theory within a pair of $(M{\bar M})$-brane \cite{kar-PRD-R}. A dynamical NP-correction does not modify an emergent metric, rather it   incorporates a torsion and hence turns out to be non-Riemannian. Thus the quintessence QFT becomes insignificant in the GTR, which is a classical theory in $d$$=$$4$.

\sp
\noindent
{\bf Bulk/Boundary in higher dimensions:}
Generically a massless NS form quantum dynamics in $(p+1)$-dimensions is completely described by a metric tensor in $p$-dimensional classical theory in presence of a HS-QFT. Furthermore the NP-theory of emergent gravity ensures a generic correspondence between a fat $(p+1)$-brane in bulk and the boundary dynamics on a gravitational $p$-brane within a vacuum created pair of $(p{\bar p})$-brane. Thus a fat $(p+1)$-brane in a strong coupling bulk gauge theory is dual to a weakly coupled boundary gravitational $p$-brane. 
The generic nature of $(bulk)_{p+1}/(boundary)_p$ correspondence in a NP-theory of emergent gravity is remarkable. It may provide a clue to an unified description of all four fundamental forces in nature.

\sp
\noindent
Interestingly the ${\rm CFT}_6/{\rm AdS}^Q_5$ correspondence may be reviewed from the perspective of $d$$=$$12$ {\it fundamental} theory \cite{kar-PRD-R}. In principle a space filling fat $9$-brane, in type IIB superstring theory, underlies a gravitational $8$-brane within a vacuum created pair of $(8{\bar 8})$-brane. Similarly a space filling boundary gravitational $9$-brane may urge for a fat $M$-brane in $d$$=$$11$ bulk and viceversa. Preliminary analysis reveals that $d$$=$$12$ (higher) form theory, $i.e.$ a $B_p$ form for all $p\ge 2$, can be a potential candidate for a {\it fundamental} bulk theory and the corresponding boundary dynamics may describe the $M$-theory in $d$$=$$11$. It provokes thought to believe that the higher form(s) theory with a self-dual $6$-form may describe all five superstring vacua in $d$$=$$10$ in an emergent quantum gravity scenario on pairs of $(9{\bar 9})$-brane. The detailed discussion is beyond the scope of this article and is in progress.

\sp

\sp
\noindent
{\it\bf Acknowledgements:}
Author (SK) acknowledges the Research and Development grant RC/2015-16/9677 by the University of Delhi, India.

\def\anp{Ann. of Phys.}
\def\cmp{Comm.Math.Phys.\ {}} {}
\def\springer{Springer.Proc.Phys.}
\def\prl{Phys. Rev. Lett.}
\def\prd#1{{Phys.Rev.} {\bf D#1}}
\def\jhep{JHEP\ {}}{}
\def\cqg{Class.\& Quant. Grav.}
\def\plb#1{{Phys. Lett.} {\bf B#1}}
\def\npb#1{{Nucl. Phys.}{\bf B#1}}
\def\mpl#1{{Mod. Phys. Lett} {\bf A#1}}
\def\ijmpa#1{{Int. J. Mod. Phys.} {\bf A#1}}
\def\ijmpd#1{{Int. J. Mod. Phys.} {\bf D#1}}
\def\mpla#1{{Mod.Phys.Lett.} {\bf A#1}}
\def\rmp#1{{Rev.Mod.Phys.} {\bf 68#1}}
\def\jmp#1{{J.Math.Phys.}}
\def\jaat{J.Astrophys.Aerosp.Technol.\ {}} {}
\def \epj#1{{Eur.Phys.J.} {\bf C#1}} 
\def \jcap{JCAP\ {}}{}

\end{document}